\documentstyle[aps,manuscript]{revtex}
\newcommand{\beq}{\begin{equation}}
\newcommand{\eeq}{\end{equation}}
\newcommand{\ba}{\begin{array}}
\newcommand{\ea}{\end{array}}
\begin{document}

\title{Ultra High Energy Neutrinos from Hidden-Sector Topological Defects}

\author{V.S. Berezinsky}
\address{INFN, Laboratori Nazionali del Gran Sasso,\\
I-67010 Assergi (AQ), Italy}
\author{A. Vilenkin}
\address{Institute of Cosmology, Department of Physics and Astronomy,\\
Tufts University, Medford, MA 02155,USA}

\maketitle
\begin{abstract}
We study Topological Defects (TD) in hidden (mirror) matter 
as possible sources of ultra-high energy neutrinos. The hidden/mirror and
ordinary matter are assumed to interact very weakly through gravity 
or superheavy particles.  An inflationary scenario is outlined in
which superheavy defects are formed in hidden/mirror  matter (and not in
ordinary matter), and at the same time the density of 
mirror matter produced at the end of inflation is much smaller than
that of ordinary matter.  
Superheavy particles produced by hidden-sector TD and the products of
their decays are all sterile in our world. Only mirror neutrinos
oscillate into ordinary neutrinos. We show that oscillations with 
maximal mixing of neutrinos from both       
worlds are possible and that values of $\Delta m^2$, needed for for
solution of solar-neutrino and atmospheric-neutrino problems, 
allow the oscillation of 
ultra-high energy neutrinos on a timescale of the age of the Universe. 
A model of mass-degenerate visible and mirror neutrinos with maximal
mixing is constructed. Constraints on UHE neutrino
fluxes are obtained. The estimated fluxes can be 3 orders of magnitude
higher than those from ordinary matter. Detection of these fluxes is 
briefly discussed. 
\end{abstract}
\section {Introduction}
  A hidden sector of mirror particles was first suggested   
by Lee and Yang \cite{LeYa} in 1956 to save the conservation of parity in the 
whole enlarged particle space. This concept was further discussed and 
developed in Ref.\cite{SaKOP}. Later the idea of two weakly
interacting sectors, visible and hidden, found interesting phenomenological 
applications and development \cite{mirror,KST}. It
has been boosted in 1980s by superstring theories with $E_8\times E'_8$
symmetry.
The particle content and symmetry of interactions  in each
of the $E_8$ groups are identical, and thus the mirror world has naturally
emerged.\\
The most recent reincarnation of hidden-sector models is in the
context of D-branes \cite{Polchinski,Dvali}.  In this approach, light
particles are associated with the endpoints of open strings which are
attached to D-branes.  Ordinary and hidden-sector particles live on
different branes which are embedded in a higher-dimensional
compactified space.

How do the ordinary and hidden sectors communicate with each other?\\
Most naturally they interact gravitationaly. This possibility is employed  
in \cite{BeMo}. More generally, and this is also discussed in
\cite{BeMo} and \cite{Bz98}, ordinary and mirror matter can also interact
through the exchange of superheavy gauge particles. 
In the D-brane context, in some models the interaction between
different branes occurs only through the exchange of closed strings
(gravitons), while other models (in which the two branes are embedded
in a brane of higher dimensionality) allow for a gauge boson mediated
interaction. 

In the case of gravitational 
coupling of the two worlds, one should expect dimension 5
gravitational interaction scaled by the Planck mass $m_{pl}$; in the case
of superheavy gauge bosons this scale might be $\Lambda < m_{pl}$: 
\beq 
{\cal L} \sim \frac{1}{\Lambda}(\psi H)(\psi' H'),
\label{d5}
\eeq
where $\psi$ and $H$  are respectively the 
lepton and Higgs SU(2) doublets, with mirror
fields denoted by primes. Eq.(\ref{d5}) 
provides mixing of ordinary and mirror neutrinos and neutrino masses 
\cite{BEG},\cite{ABS}.  
Because of the smallness of the neutrino masses, this is the 
most visible physical
effect caused by the gravitational interaction ($\Lambda=m_{pl}$) of
particles from the two worlds.
 
In principle, there could be other ways of communication. For example,
(see \cite{FoVo}) one can add to the Lagrangian a Higgs potential term 
$\lambda \phi \phi\phi'\phi'$ and a gauge boson kinetic mixing
term $hF_{\mu\nu}F'_{\mu\nu}$, with $\lambda$ and $h$ being new
coupling constants. These terms have potential problems. 

The discrete $P$-symmetry that interchanges the two worlds can be
spontaneously broken.  In this case, the coupling constants, the Higgs
potentials and expectation values, and even the symmetry breaking
patterns will generally differ from one world to the other.  The
breaking of $P$-symmetry can be implemented by giving a non-zero vev
to a spin-0 field which transforms as a singlet under the gauge groups
in both sectors ans as a pseudoscalar under the $P$-transformation:
$P\phi=-\phi$.  Models of
this type have been studied in Refs.\cite{BeMo,BDM}; we shall refer to
them as {\em asymmetric} hidden sector models.\footnote{One can, of course,
consider models without $P$-symmetry, in which the ordinary and mirror
sectors are not symmetric to begin with.  For our purposes, such
models are essentially the same as models with a spontaneously broken
$P$-symmetry.} 

A model with an unbroken $P$-symmetry has been developed in
Ref.\cite{FoVo}.  The $P$-transformation in this model turns the 
left-handed ordinary fermions into right-handed mirror fermions. 
The masses and couplings of ordinary and mirror particles are
identical, and hence the term ``mirror matter'' is more justified in
this case.  The 
EW Higgs fields in both sectors are also parity partners and have
equal vev's.  Mixing of neutrinos from different
worlds (taken as an ad hoc
terms in the Lagrangian) provides  a channel of 
communication between the worlds:
ordinary neutrinos can oscillate into mirror neutrinos, which play the
role of sterile neutrinos. We shall refer to models of this type (with
model \cite{FoVo} as the most elaborated example) as 
{\em symmetric} hidden sector model.

Although the term ``mirror sector'' suggests that it is related to the
ordinary sector by a reflection transformation, most of the discussion
in this paper is applicable to a more general class of hidden-sector
models.  In what follows, we shall use the terms ``mirror sector'' and
``hidden sector'' interchangeably.
 
Mirror neutrinos are the most natural candidates for sterile neutrinos
often considered now for explaining the oscillations of solar and
atmospheric neutrinos. In this respect mirror neutrinos have been studied
in Refs.\cite{BeMo,FoVo}.

Mirror matter has cosmological consequences which result in 
model restrictions.

In symmetric hidden sector models, 
the number of massless and light particles is 
doubled in comparison with ordinary matter, and this case is excluded
by cosmological nucleosynthesis, if the temperatures of mirror and
ordinary matter are the same. (More generally, in any mirror model 
the effective number of light degrees of freedom is larger than
for the ordinary matter, and this number is restricted by nucleosynthesis.) 

One way to suppress the light degrees of freedom is to diminish the
temperature of the mirror matter in the Universe \cite{FoVo,BDM} (see 
further discussion in Section II). This reduces the number density of mirror
photons in a sraightforward way, while the situation with sterile neutrinos
is more delicate \cite{FoVoPRL}. Even if the initial density of
mirror neutrinos is negligible, they reappear again and may be
brought to equilibrium at nucleosynthesis epoch due to oscillations 
between ordinary and mirror neutrinos \cite{reapp}. Nucleosynthesis 
constraint bounds the allowed neutrino properties in the parameter space 
($\Delta m^2 \;,\; \sin^22\theta$), where $\theta$ is the mixing
angle \cite{BBN}. It is clear that the larger the mixing angle is, the
smaller are the allowed values of $\Delta m^2$.  Electron neutrinos impose
the strongest limit on $\Delta m^2$, because 
$\nu' \leftrightarrow \nu_e$ oscillation influences nucleosynthesis
not only through the rate of the cosmological 
expansion, but also due to the rate
of $n \leftrightarrow p$ conversion. 

A cruicial assumption involved in deriving the bounds described above
is that the relic lepton asymmetry is absent. In the presense of a large
lepton asymmetry, $L \gtrsim 5\cdot 10^{-5}$, the potential for
active neutrinos results in a small mixing angle between active and
sterile (mirror) neutrinos, and thus in a weak oscillation 
between these components at temperatures relevant for nucleosynthesis 
\cite{FoVoPRL},\cite{Vo},\cite{KiCh}. Therefore, in this case  
$\nu' \leftrightarrow \nu$ oscillations and induced by them the
additional number of degrees of freedom are suppressed.
\footnote{In the scenario studied in this paper only 
muon neutrinos from ordinary and mirror worlds are maximally mixed, while the 
other two types of neutrinos might have very small mixing. Therefore,
even in case $L=0$, this 
scenario gives only one extra neutrino and thus marginally survives the 
nucleosynthesis restrictions}.  

A cosmological origin for the temperature difference between mirror and
ordinary matter in the universe has been already considered in
Refs.\cite{KST},\cite{BDM}. The asymmetry is generated as a result of
different rates of inflaton decay to ordinary and mirror matter due 
to $P$-symmetry breaking. 



In Ref.\cite{BDM} mirror neutrinos and 
baryons were considered as dark matter particles. The problem of structure 
formation
with mirror matter and other astrophysical implications were studied 
in Refs.\cite{BDM},\cite{Be},\cite{MoTe}.  

In this paper we shall study mirror matter in the Universe as 
a source of ultra-high energy neutrinos. As concrete sources we
shall consider mirror Topological Defects. They produce high 
energy mirror neutrinos in the usual way: through production and decay of
superheavy mirror X-particles. Then high
energy mirror neutrinos oscillate into ordinary neutrinos, while the 
other products of decay of mirror X-particles remain in the mirror world,
being invisible in the ordinary matter. These sources give an ideal 
example of ``hidden neutrino sources'' \cite{book}. High energy
neutrino radiation from ordinary sources is usually accompanied by 
other radiations, most notably by high energy gamma rays. Even in cases when 
high energy photons are absorbed in the source, their energy is partly
transformed into low energy photon radiation: X-rays or thermal radiation. 
The fluxes of these radiations impose an upper bound on the high energy
neutrino flux. For sources transparent for HE gamma radiation, in 
particular for Topological Defects (TD), the upper limit on diffuse 
neutrino flux is imposed by the cascade e-m radiation
\cite{BeSm},\cite{book}. In all cases (e.g. decay of X-particles, $pp$ and
$p\gamma$ interaction) neutrinos and photons are produced at the decays of
pions. Colliding with microwave photons, high energy photons and electrons
produce e-m cascade with most of the 
energy being in the observed $100~MeV - 10~GeV$
energy range. The energy density of this cascade radiation should not exceed,
 according to EGRET observations, $\omega_{cas} \sim 1-2\cdot 10^{-6}~eV/cm^3$.
Introducing the neutrino energy density for neutrinos with individual energies 
higher than E, $\omega_{\nu}(>E)$, it is easy to derive the following chain
of inequalities (from left to right):
\beq
\omega_{cas}>\omega_{\nu}(>E)=\frac{4\pi}{c}\int_E^{\infty}EI_{\nu}(E)dE>
\frac{4\pi}{c}E\int_E^{\infty}I_{\nu}(E)dE=\frac{4\pi}{c}EI_{\nu}(>E),
\label{cas}
\eeq
An upper bound on the integral diffuse neutrino
flux immediately follows From Eq.(\ref{cas}):
\beq
I_{\nu}(>E) < \frac{c}{4\pi}\frac{\omega_{cas}}{E}.
\label{lim}
\eeq

The upper bound given by Eq.(\ref{lim}) is the most restrictive one for 
diffuse neutrino fluxes produced by ``ordinary'' TD, AGN, 
gamma-ray bursts, etc. 
The neutrino flux from hidden-sector TD is free from this bound, though in
Section V  
we shall present another (weaker) upper limit on high-energy neutrino
flux from hidden-sector TD. 

Another problem where UHE neutrinos from mirror TD can be helpful is
the production of Ultra High Energy Cosmic Rays (UHECR) by resonant neutrinos.

The signature of extragalactic UHECR, the Greisen-Zatsepin-Kuzmin (GZK)
cutoff \cite{GZK} is not found in observations \cite{UHECR}.
Three ideas have been suggested to explain the absence of the cutoff: 
(i) signal carriers are not absorbed
on the microwave radiation, (ii) the sources form a compact group around our
Galaxy with a
linear size smaller than the GZK absorption lenght, and (iii) the sources
are distributed uniformly in the Universe, but the target particles on which
UHECR are produced by the signal carriers form a compact object. 

The case (iii) is realized \cite{WeFa} with the help of UHE neutrinos (signal
carriers) interacting with relic neutrinos which have an enhanced density 
in some Large Scale Structure near us. One of the problems with this idea is
that it is not clear how one can arrange a large flux of UHE
neutrinos.  The origin of this flux is left unspecified
in all publications ({\em e.g.} \cite{Sigl}, \cite{GeKu}), \cite{Zas}  that 
we are aware of. The very high resonant neutrino energy,   
$E_0 \sim 4\cdot 10^{12}~GeV$ for $m_{\nu} \sim 1~eV$, implies a top-down 
scenario, but the fluxes of neutrinos in such scenarios are
limited by the cascade constraints and are too small to produce the observed
flux of UHECR. Mirror TDs, which evade the cascade constraints, can in
principle produce the desirable UHE neutrino flux. This problem is
addressed in Section VI.

The ouline of this paper is as follows.

In Section II we review inflationary scenarios for the mirror
Universe, elaborating in particular a two-inflaton scenario for
symmetric HS models. Formation of hidden-sector TD is discussed in
Section III.  Fluxes of High Energy
Neutrinos from TD are calculated in Section IV for the case of
necklaces as an example. An upper bound on the neutrino flux from
hidden-sector 
TD is obtained in Section V. UHECR from resonant neutrinos are studied
in Section VI. The main results of this paper are summarized in
Section VII. 
In Appendix our model for mass-degenerate neutrinos with maximal mixing
is described. 

Throughout the paper we shall use following abbreviations: HE and UHE for
High Energy and Ultra High Energy, respectivey, UHECR - for Ultra High
Energy Cosmic Rays, TD - for Topological Defects, HS - for Hidden
Sector (including mirror sector), LSS - for a 
Large Scale Structure in the universe, LG - for the Local Group of galaxies,
LS -for the Local Supercluster of galaxies.

\section{Inflationary scenarios for hidden-sector Universe}


As was discussed in the Introduction, there are several dangers to be
watched for in models with mirror
matter.  The main one is a possible conflict with the standard
nucleosynthesis.  If the mirror sector contains a massless photon and
three light neutrinos, and the temperatures of the two worlds are the
same, then the density of mirror matter at the time of
nucleosynthesis is unacceptably high (it amounts to five extra
neutrino species).   
 
Two conditions are necessary to overcome this problem (see Introduction):
The mirror matter must have a lower temperature $T'\lesssim 0.5 T$ and 
lepton asymmetry is needed to suppress excessive production of mirror 
neutrinos through oscillation of ordinary neutrinos.  
Here, we
shall discuss some inflationary scenarios that can naturally lead to a
temperature difference between the two worlds.

The condition $T'\lesssim 0.5 T$ can be easily satisfied in asymmetric
hidden sector models.  For example, we could have a single inflaton
field $\phi$ which transforms as a scalar (or pseudoscalar
\cite{BeMo}) under the $P$-transformation.  Since the symmetry between
the two sectors is broken, the field $\phi$ will generally have
different couplings to particles in different sectors.  It will then
decay into ordinary and mirror particles at different rates, and the
reheating temperature in the mirror matter can be lower \cite{KST,BDM}.
  
This scenario would not work in symmetric hidden sector models: 
the inflaton would then have identical couplings to both sectors, and
the two reheating temperatures would be the same.
This problem can be addressed in the following two-inflaton scenario
(see also \cite{KST}). 


Let us consider two inflaton fields, $\phi$ and
$\phi'$, with $\phi$ belonging to the visible sector and $\phi'$ to 
the mirror sector.  During inflation,
both inflatons roll down towards the minima of their respective
potentials.  Inflation ends when both of them have reached their
minima.  An important point is that the evolution of $\phi$ and $\phi'$
need not be synchronized.  The inflaton dynamics is influenced by
quantum fluctuations which cause inflation to end at different times
in different regions of space \cite{inflation}.  In our case there are
two inflatons, their fluctuations are uncorrelated, and 
one expects them to reach their minima at different
times, even in the same spatial regions.  In regions where $\phi'$
reaches minimum first, any mirror particles produced due to its
oscillations are diluted by the remaining inflation driven by the field
$\phi$.  By the time when the energy of $\phi$ thermalizes, the
density of mirror matter will then be very small, so that $T\gg T'$.
Note that the (co-moving) coherence length of the inflaton fields
should be much greater than the present horizon, so we expect very
large (super-horizon) regions of the universe to be dominated by
ordinary matter, and similarly large regions dominated by mirror
matter, with relatively tiny boundary regions where both kinds of
matter are present in comparable amounts.  It is very unlikely for us
to find ourselves in one of such rare regions.

A quantitative analysis shows that this de-synchronization picture
may or may not apply, depending on the form of the inflaton potential 
$U(\phi)$.  We shall see, however, that it does apply for the simplest
choice of the potential,
\beq
U(\phi)=m_\phi^2\phi^2/2.  
\label{Um}
\eeq

In ``chaotic'' inflation scenario, the inflatons roll from very large
values of $\phi$ towards $\phi=0$. The initial values of $\phi$ and 
$\phi'$ are large and uncorrelated. At very large $\phi$, $\phi > \phi_q$,
the dynamics of $\phi$ are dominated by quantum fluctuations. The
boundary $\phi_q$ of this ``quantum diffusion'' regime is 
determined by condition $dU/d\phi \sim H^3$. 
At  
$\phi_*\lesssim\phi\lesssim\phi_q$, where $\phi_* \sim m_{pl}, \;\; \phi$
(and $\phi'$) evolve in a slow-roll regime described by the equations 

\beq
3H{\dot \phi}=-\frac{dU(\phi)}{d\phi},
\label{hphi}
\eeq
\beq
3H{\dot {\phi'}}=-\frac{dU(\phi')}{d\phi'},
\label{hphi'}
\eeq
\beq
H^2={8\pi G\over{3}}[U(\phi)+U(\phi')],
\label{H2}
\eeq

Will the ``incidental'' initial ratio $\phi'/\phi$ be conserved during
the slow-roll evolution?
For the potential (\ref{Um}) the answer is ``yes''. Indeed, the integration
of Eqs. (\ref{hphi}) - (\ref{H2}) results in 
\beq
\phi'/\phi=const.
\eeq  

Suppose that $\phi'\ll\phi$ at the time when $\phi$ begins its slow
roll ($\phi\sim\phi_q\sim m_{pl}^{3/2}m_\phi^{-1/2}$), 
so that we can neglect $U(\phi')$ in
(\ref{H2}).  The characteristic value of $\phi'$ is then determined by
quantum fluctuations about $\phi'=0$ and can be found from \cite{SY}
$U(\phi')\sim H_q^4$, where $H_q\sim (m_\phi m_{pl})^{1/2}$ is
the expansion rate at $\phi\sim\phi_q$.  This gives $\phi'\sim m_{pl}$
and $\phi'/\phi\sim (m_\phi/m_{pl})^{1/2}$.  This small ratio of
$\phi'/\phi$ is preserved all the way to the end of inflation.
In this type of models,
the universe is divided into super-horizon regions dominated by
ordinary matter and equally large regions dominated by mirror matter.

For a different choice of the potential, e.g. 
\beq
U(\phi)=\lambda_\phi m_{pl}^4(\phi/m_{pl})^n,
\label{Uphi}
\eeq
with $n>2$,
the fields $\phi$ and $\phi'$ do get synchronized at 
late stages of the evolution.  (For $n=4$, this was shown in
Ref. \cite{KL}.)  In this case, 
integration of the slow-roll equations (\ref{hphi}) -(\ref{H2}) gives
\beq
\phi^{-n+2}-{\phi'}^{-n+2}=const.
\label{solphi}
\eeq
At the onset of the slow roll of $\phi$, when $\phi\sim\phi_q\sim
\lambda^{-1/4}m_{pl}$, the typical value of $\phi'$ is $\phi'\sim
\lambda_\phi^{(2-n)/n(n+2)}m_{pl}$.  With $\lambda_\phi\ll 1$, this is
much smaller than $\phi_q$ but still much larger than $\phi_*$.  
Now, consider the solution (\ref{solphi}) with these initial values of
$\phi$ and $\phi'$.  By the end of inflation, both fields get much
smaller than their initial values, so that the  constant on the
right-hand side of (\ref{solphi}) becomes unimportant, and we have
$\phi'\approx\phi$.  
Thus, $\phi$ and $\phi'$ get synchronized by the
end of inflation, even if they were not initially.  We conclude that
models with a power-law potential (\ref{Uphi}) and $n>2$ give equal
densities of mirror and ordinary matter.

Coming back to our basic ordinary-matter dominated scenario, we can give
another example of a model with segregated mirror and ordinary matter
is a two-inflaton model where inflation occurs at a metastable minimum
of the inflaton potential.  The highest rate of inflation is achieved
in the false vacuum state where 
both $\phi$ and $\phi'$ are at the minima of their respective
potentials.  This state decays through nucleation of two types of
bubbles.  In bubbles of the first type, the field $\phi'$ tunnels
through the barrier and starts rolling down its potential, while
$\phi$ remains in the false vacuum.  As $\phi'$ rolls to the bottom of
the potential and decays into mirror particles, inflation continues in
the interior of the bubble.  Mirror particles are quickly diluted
away, and the bubble interior is filled with inflating false vacuum of
the ``ordinary'' inflaton $\phi$.  This vacuum will in turn decay
through nucleation of bubbles of the second type (with $\phi$
tunneling and $\phi'$ remaining unchanged).  If the potential is
sufficiently flat, the roll down of $\phi$ is accompanied by
additional inflation, and after the $\phi$-field decay, the interiors
of these secondary bubbles will become dominated by ordinary matter.
Quite similarly, nucleation of $\phi'$-bubbles inside $\phi$-bubbles
results in regions dominated by mirror matter.

\section{Hidden-Sector Topological Defects}

Apart from the general considerations which apply to any model with a
hidden sector (HS), we have to address some additional issues specific to
HS defects.  First we have to arrange for these defects to
form.   And second, we have to avoid the formation of similar defects
in our sector, since otherwise we would get unacceptably large fluxes
of ordinary UHE particles, and the cascade bound would be violated.
 
Once again, these conditions can be easily satisfied in asymmetric HS
models, where the symmetry breaking scales and even the symmetry
breaking patterns may be different in the two worlds.
Perhaps the simplest possibility is the model of Ref.(\cite{BDM}) with one
inflaton and asymmetric reheating.
HS defects can be formed
in a usual symmetry-breaking phase transition after inflation.  We only
have to arrange for the corresponding phase transition in the ordinary
matter to occur at a lower energy scale or not to occur at all.
\footnote{In
models discussed in Refs. \cite{BeMo,BDM}, the $P$-symmetry breaking
is assumed to be at the electroweak scale, $\eta_P\sim\eta_{EW}\sim
10^2 GeV$, and the two worlds have nearly identical physics 
above $\eta_{EW}$.  But this needs not be the case.  The physics of
mirror and ordinary defects will be different in models where $\eta_P$
is greater than the symmetry breaking scale of the defects.}

In symmetric HS models, the two sectors have identical physics and
identical defect solutions.  It does not follow, however, that
ordinary and HS defects should be present in the universe in equal
numbers.  The density of defects is determined by the cosmological
evolution, which can be different for the two sectors.  For example,
in the two-inflaton model discussed above, mirror matter is completely
inflated away, and if defects were formed in phase transitions after
inflation, we would expect to have ordinary defects but no HS
defects.  For our purposes, however, we need the opposite situation: HS
defects and no ordinary defects.

  This can be arranged if defects are
formed in a curvature-driven phase transition during inflation.

As an illustration we shall consider a toy model of a spontaneously
broken $SU(2)$ symmetry.  We introduce a Higgs triplet ${\bf
\chi}=(\chi_1,\chi_2,\chi_3)$ with a potential
\beq
V({\bf \chi})={1\over{4}}\lambda(|{\bf\chi}|^2-\eta^2)^2.
\eeq
When ${\bf\chi}$ acquires an expectation value, $SU(2)$ is broken to
$U(1)$ and monopoles are formed.  Suppose now that ${\bf\chi}$ is
coupled to the inflaton $\phi$,
\beq
V_\phi({\bf\chi})=-{1\over{2}}g\phi^2|{\bf\chi}|^2,
\eeq
and has a non-minimal coupling to spacetime curvature,
\beq
V_R({\bf\chi})={1\over{2}}\xi R|{\bf\chi}|^2.
\eeq
The mirror field ${\bf\chi}'$ has identical couplings to the mirror
inflaton $\phi'$ and to the curvature.

We shall assume ``chaotic'' inflation with a quadratic inflaton
potential (\ref{Um}).  After the mirror inflaton $\phi'$ reached its
minimum, the effective mass of the mirror field ${\bf\chi'}$ is
\beq
m_{\chi'}^2=-\lambda \eta^2 +\xi R,
\eeq
and the curvature is given by
\beq
R\approx 12H^2\approx 16\pi (m_\phi/m_{pl})^2\phi^2.
\eeq
We see that  above  the
critical curvature
\beq
R_{\chi'}=\lambda\eta^2/\xi.
\eeq
the expectation value of ${\bf\chi'}$ vanishes and the $SU(2)'$
symmetry is restored.
Thus, even in the absence of mirror matter, as the curvature gradually
decreases during inflation, we can have symmetry-breaking phase
transitions accompanied by the formation of HS defects.  If these
curvature-driven phase transitions occur sufficiently close to the end
of inflation, the defects will not be completely inflated away and can
serve as sources of UHE neutrinos.

The effective mass of the field ${\bf\chi}$ in the ordinary sector is
\beq
m_\chi^2=-\lambda\eta^2 -g\phi^2 +\xi R
=-\lambda\eta^2 -g[1-16\pi(\xi/g)
(m_\phi/m_{pl})^2] \phi^2.
\label{mchi}
\eeq
We see that $m_\chi^2<0$ for all $\phi$, provided that
\beq
16\pi(\xi/g)(m_\phi/m_{pl})^2<1.
\eeq
In this case no ordinary defects are formed during the whole slow roll
period of $\phi$ (and any defects formed prior to that are completely
diluted away)

One final condition that has to be checked is that the temperature at
reheating is not so high that the symmetry gets restored, since
otherwise ``ordinary'' defects will be formed again in a subsequent
phase transition.  All the above conditions can be satisfied without
fine-tuning.

%
%

\section{ Fluxes of high-energy neutrinos from HS defects}

We shall consider necklaces \cite{BeVi} as a specific example of TD. 
Necklaces are hybrid TDs formed by monopoles (M) and 
antimonopoles ($\bar{M}$), each being attached to two strings. The monopole 
mass $m$ and the mass per unit length of string $\mu$ are determined by 
the corresponding 
symmetry breaking scales , $\eta_s$ and $\eta_m$,
\beq
m \sim 4\pi \eta_m/e, ~~~~~~ \mu \sim 2\pi \eta_s^2
\label{vev}
\eeq
where $e$ is the gauge coupling.  The evolution of necklaces depends
on the parameter
\beq
r=m/\mu d
\eeq
which gives the 
ratio of the monopole mass to the average mass of string between two monopoles
($d$ is the average string length between monopoles).  It cannot
exceed $r_{max}\sim\eta_m/\eta_s$.
As it is argued in ref.\cite{BeVi}, necklaces might evolve towards a
scaling solution
with a constant $r\gg 1$, possibly approaching $r\sim r_{max}$.
Monopoles and antimonopoles trapped in the necklaces inevitably
annihilate in the end, producing 
superheavy Higgs and gauge bosons (X particles) of mass
$m_X \sim e\eta_m$. The rate 
of $X$-particle production per unit volume and time is
\beq
\dot{n}_X \sim r^2 \mu/t^3m_X
\label{rate}
\eeq

 From the relations above it is easy to see that 
\beq
\zeta \equiv \frac{r^2\mu}{m_X^2}=\frac{2\pi}{e^2}\left (\frac{r}{r_{max}}
\right )^2 \lesssim 10.
\label{zeta}
\eeq

High energy neutrinos are produced in the chain of X-decays via pions. 
For simplicity we assume that pions (of all charges) are produced with 
a power-law spectrum
\beq
D_{\pi}(x, m_X) = 4(2-p)2^{-p} x^{-p}
\label{pions}
\eeq
where $x=E_{\pi}/m_X$ is a fraction of energy taken away by a pion, and for 
$p$ we shall use a value between 1.3 and 1.5 , which bound the realistic 
QCD spectrum of pions. The 
spectrum (\ref{pions}) is normalized so that 
$\int_0^{1/2}x D_{\pi}(x,m_X) dx=1$. 

The number of neutrinos with energy $E_{\nu}$ from the decay of one X-particle 
is given by
\beq
N(E_{\nu})=\frac{4}{m_X}\int_{2E_{\nu}/m_X}^{1/2} \frac{dx}{x}D_{\pi}(x,m_X).
\label{nu-numb}
\eeq
The diffuse flux of mirror neutrinos $\nu'_i$ (where $i=e$ and $\mu$ , 
antineutrinos are not included) is

\beq
I_{\nu_i}(E)=\frac{c}{\pi}\frac{\dot{n}_X(t_0)}{m_X H_0}
\int_0^{\infty} dz \int_{2E_{\nu}(1+z)/m_X}^{1/2}\frac{dx}{x}D_{\pi}(x,m_X)
\label{diff}
\eeq
 Finally, we obtain the diffuse flux of ordinary neutrinos $\nu_i$ taking into 
account $\nu'_i \to \nu_i$ oscillation with averaged probability 
$P_{osc}\sim 1/2$:
\beq
I_{\nu_i}(E)=\frac{c}{4\pi}\frac{\dot{n}_X(t_0)}{m_X H_0}
\left ( \frac{E_{\nu}}{m_X/2}\right )^{-p}\frac{(2-p)2^{2-p}}{p(p-1)}P_{osc}
\label{nu-diff}
\eeq

This expression can be written in more compact form:
\beq
I_{\nu_i}=k_p \zeta \frac{c}{4\pi}\frac{1}{t_0^2}
\left ( \frac{E_{\nu}}{m_X}\right )^{-p},
\label{nu-flux}
\eeq
with $\zeta$ given by Eq.(\ref{zeta}) and 
\beq
k_p=\frac{6(2-p)2^{-2p}}{p(p-1)}P_{osc}.
\eeq
For $P_{osc}=1/2$ and $p=1.5 ~~ k_p=1/4$ and $k_p=0.89$ for $p=1.3$. 

The neutrino flux in Eq.(\ref{nu-flux}) can be very large. 
For example, with $r^2\mu \sim 0.1m_X^2, \;\; m_X \sim 
10^{16}~GeV$, $p=1.5$ and $P_{osc}=1/2$, one obtains at $E \sim 10^{11}~GeV$, 
$E^3 I_{\nu} \sim 1\cdot 10^{28}~eV^2m^{-2}s^{-1}sr^{-1}$, i.e. a flux
three orders 
of magnitude larger than predicted from ordinary sources under most
optimistic assumptions. 
Since we have used 
a very rough power-law approximation for the spectrum, it may be better to 
illustrate the enhancement of the flux by comparing Eq.
(\ref{nu-diff}) to the maximum allowed flux from ordinary sources 
with the same power-law spectrum. 

Suppose that some unidentified ordinary-matter sources produce 
superheavy X-particles, which decay producing high energy pions. 
Then the energy density of neutrinos $\omega_{\nu}$ and the cascade density 
$\omega_{cas}$ are equal. Using this fact we immediately obtain an upper 
limit on the neutrino flux as 
\beq
I_{\nu_i}(E) \leq (2-p)2^{-p}\frac{\omega_{cas}}{m_X^2}\frac{c}{4\pi}
\left( \frac{E}{m_X} \right )^{-p}
\label{nu-ord}
\eeq
 The ratio of the two fluxes [Eq.(\ref{nu-diff}) and Eq.(\ref{nu-ord})] 
is given basically by the value 
$m_X^2/(\omega_{cas} t_0^2)$ and it is $\sim 2\cdot 10^4$ 
for $m_X \sim 1\cdot 10^{16}~GeV$ and for 
observationally allowed cascade density 
$\omega_{cas} \sim 2\cdot 10^{-6}~eV/cm^3$.

\section{E-m cascade restrictions}

All particles from the QCD cascades produced by
decays of mirror X-particles are sterile in our world.
Only mirror neutrinos can oscillate into ordinary ones.
 An upper bound on the neutrino flux is 
given by the resonant interaction of UHE neutrinos with relic cosmological 
neutrinos, $\nu +\bar{\nu} \to Z^0 \to pions$. Electrons and 
photons from the decay of pions initiate e-m cascades on the 
microwave radiation. Reactions $\nu +\bar{\nu} \to Z^0 \to l+\bar{l}$,
where $l=e, \mu, \tau$, also contribute to the cascade.
The calculated cascade energy density $\omega_{cas}$ must be smaller than 
the energy density $\omega_{\gamma}^{obs}$ observed ({\em e.g.} by EGRET) in 
the extragalactic diffuse radiation.  

Let us first derive a convenient formula for the rate of resonant events.

The resonant neutrino energy $E_0$ and the resonant cross-section 
$\sigma(E)$ for $\nu+ \bar{\nu} \to Z^0 \to f$ ($f$ is an arbitrary final
state) are given by 
\beq
E_0=\frac{m_Z^2}{2m_{\nu}}=4.16\cdot 10^{12}\left (\frac{1~eV}{m_{\nu}}\right )
~GeV
\label{res}
\eeq
\beq
\sigma_{\nu,f}(E_c)=\frac{12\pi}{m_Z^2}\frac{\Gamma_{\nu}\Gamma_f}
{(E_c-m_Z)^2+\Gamma_t^2/4}
\label{c-sect}
\eeq
Here, $m_Z$ is the mass of $Z^0$-boson, $E_c$ is the center-of-mass 
energy, $\Gamma_{\nu},\;\;
\Gamma_f$ and $\Gamma_t$ are the widths of $Z^0$ decay to 
neutrinos, to an arbitrary final state $f$ 
and the total width, respectively. In Eq.(\ref{c-sect}) we 
took into account that only one chiral component of neutrino takes part in 
the interaction. \footnote{Production of $Z^0$ occurs through the interaction
of flavor neutrinos, {\em e.g.} $\nu_{eL}+\bar{\nu}_{eL}$ in the case of
Dirac neutrinos, or $\nu_{eL}+\nu_{eL}^c$ in the case of Majorana neutrinos. In
practice one considers interaction of HE flavor neutrino, 
{\em e.g.},  $\bar{\nu}_e$, with a 
physical mass-eigenstate target neutrino $\nu_1$
of mass $m_1$. The probability to find this neutrino as $\nu_e$ is 
equal to $\cos^2\theta$ (or $\sin^2\theta$), where $\theta$ is the mixing 
angle. Therefore, the cross-section in Eq.(\ref{c-sect}) must include 
$\xi=\cos^2\theta$ (or $\sin^2\theta$). In the case of mass-degenerate 
neutrinos, 
$m_1 \approx m_2 =m_{\nu}$, an incident HE $\bar{\nu}_e$ interacts with 
$\nu_e$ component of $\nu_2$ almost at the same resonant energy and 
$\xi=1$. 
There is no difference in the rates for Dirac and Majorana neutrinos:
this becomes obvious if in counting the number of neutrino species one
includes
both $\nu_L$ and $\nu_L^c$ in the case of Majorana neutrinos.}

The rate of $Z^0$ production per unit volume due to collisions of high
energy flavor neutrino $\nu_i$ with target antineutrino $\bar{\nu}_i$ 
is given by 
\beq
\dot{n}_Z= 4\pi n_{\bar{\nu}_i}\xi\int I_{\nu_i}(E)\sigma(E)dE=
4\pi \sigma_t n_{\bar{\nu}_i}\xi I_{\nu_i}(E_0) E_0,
\label{Z-rate}
\eeq
where
\beq
\sigma_t = 48 \pi f_{\nu}G_F=1.29\cdot 10^{-32}~cm^2,
\label{eff}
\eeq
$G_F$ is the Fermi constant, and here and below 
$f_s=\Gamma_s/G_F m_Z^3$, with $\Gamma_s$ being the width of the channel.
Numerically, 
$f_{\nu}=0.019$, $f_h=0.197$ and $f_{tot}=0.283$. The case of
HE $\bar{\nu}_i$ can be trivially added. Summation over $i$ takes
place in the case of mass-degenerate neutrinos. For the 
target neutrino density 
we shall use one helicity density with zero chemical potential, 
$n_{\nu_i}=56~cm^{-3}$, corresponding to the 
temperature $T=2.73 K$ of the microwave
radiation, and $\xi=\cos^2\theta,\;\;\;\sin^2\theta$ or 1, as explained in the 
footnote.

Note that Eqs.(\ref{Z-rate}),(\ref{eff}) are exact formulae because
integration in Eq.(\ref{Z-rate}) takes place over a narrow resonant peak.

The rate of $Z^0$ production with subsequent decay of $Z^0$ to an arbitrary 
channel $f$ is given by 
$\dot{n}_Z(Z\to f)=n_Zf_f/f_{tot}$. Taking into account only dominant 
hadron channels, with $f_h/f_{tot}=0.696$, and using the fact that pions 
transfer to electro-magnetic cascade half of their energy, one can find the 
energy density of electromagnetic cascade as   

\beq
\omega_{cas}=\frac{1}{2}\frac{f_h}{f_{tot}}E_0\dot{n}_Z\xi t_0=
2\pi\xi \sigma_t n_{\nu_i} t_0 I_{\nu_i}(E_0)E_0^2\frac{f_h}{f_{tot}}. 
\label{omega}
\eeq
Using $\omega_{cas} \leq \omega_{\gamma}^{obs}$ we obtain an upper limit
\beq
I_{\nu_i}(E_0) \leq \frac{2\omega_{\gamma}^{obs}m_{\nu}^2}{\pi \sigma_t
n_{\nu_i} t_0 m_Z^4\xi}\frac{f_{tot}}{f_h}
\label{lim2}
\eeq

One can see from Eq.(\ref{lim2}) that unless the production spectrum of 
neutrinos has a cutoff at some energy lower than $E_0$, the upper limit is 
provided by the lightest relic neutrino $\nu_i$. This is not 
surprising: in QCD spectra most of the 
energy is concentrated in particles of the 
highest energies. The lower $m_{\nu}$ is, the higher is $E_0$ and the more 
energy it tranferred to the e-m cascade.

The scale of neutrino masses suggested by oscillation 
solutions to the atmospheric neutrino and solar neutrino 
problems  is $m \sim \sqrt{\Delta m^2}$. This gives the following masses:
$m_{\nu}\approx 5\cdot 10^{-2}~eV$ from atmospheric neutrino oscillations, 
$m_{\nu}\approx 2\cdot 10^{-3}~eV$ from SMA MSW, 
$m_{\nu}\approx 4\cdot 10^{-3}~eV$ from LMA MSW and 
$m_{\nu}\approx 9\cdot 10^{-6}~eV$ from VO solution. The latter mass
requires a value of $m_X$ which is too large, so we disregard this case. 

It is easy to verify that  we did not 
exceed the e-m cascade limit in the calculations of section IV.
Indeed, Eqs.(\ref{nu-diff}) and (\ref{omega})
with $r^2\mu/m_X^2=0.1,\;\;\; m_X=1\cdot 10^{16}~GeV,\;\;\; 
n_{\nu_i}=56~cm^{-3}$, with $E_0$ given by Eq.(\ref{res}) and  
with  both channels $\nu_i+\bar{\nu}_i$ and $\bar{\nu_i}+\nu_i$ 
taken into account, result in 
$\omega_{cas}=1.2\cdot 10^{-7}/\sqrt{m_1}~eV/cm^3$, where 
$m_1=m_{\nu}/1~eV$. The energy density $\omega_{cas}$ is well below the 
allowed limit for $m_1=1$ and is marginally below it for 
$m_1=3\cdot 10^{-3}$.

\section{UHECR from resonant neutrinos}

Here we shall estimate the Ultra High Energy Cosmic Ray (UHECR) fluxes
produced by neutrinos from hidden-sector
TD. If the target neutrino density is enhanced in nearby
large-scale structures (LSS), such as the halo (h) of our Galaxy, the 
Local Group
(LG) and the Local Supercluster (LS), the large flux of the observed UHECR
could be generated there. Photon fluxes dominate in the production
spectra. The ratio of photon fluxes from a large-scale
structure, $I_{\gamma}^{LSS}(E)$, and from extragalactic space,
$I_{\gamma}^{extr}(E)$, can be expressed in terms of the overdensity of
target neutrinos 
in the structure, $\delta_{LSS}$, and the length of gamma-ray absorption
in extragalactic space, $R_{\gamma}(E)$, as 
\beq
I_{\gamma}^{LSS}(E)/I_{\gamma}^{extr}(E)=\delta_{LSS}R_{LSS}/R_{\gamma}(E),
\label{lss-ratio}
\eeq
where $R_{LSS}$ is the linear size of the large-scale structure. 

The overdensity factors for the galactic halo (h), Local Group (LG),
and Local Supercluster (LS) are discussed in the accompanying paper
\cite{BBV}, and for non-degenerate neutrinos they are estimated as  
\beq
\delta_\nu^{h}< 37 m_1^3,~~~~~~~~~ \delta_\nu^{LG}< 13 m_1^3.
~~~~~~~\delta_\nu^{LS} \sim 1.
\label{odens}
\eeq
Here, $m_1=m_{\nu}/1~eV$ is the 
neutrino mass in units of $1~eV$ and the overdensity is
defined as the ratio of the neutrino density with flavor i
(antineutrinos are not included) in the structure to the average density 
of the same neutrinos in the Universe, $n_{\nu_i}= 56~cm^{-3}$.
From Eqs.(\ref{lss-ratio}),(\ref{odens}) one can see that while LS does
not give an enhancement of UHECR flux, both the galactic halo ($R_h \sim
100~kpc$), and the Local Group ($R_{LG} \sim 1~Mpc$) give an
enhancement of order $(0.3 - 1)m_1^3$. Note that this excess flux
arrives without absorption. Estimates for both structures are
given as upper limits, with the limit for the galactic halo being more
reliable. In the estimates below, the index $LSS$ 
refers to one of these two structures. 

Once again, let us take HS necklaces as an 
example of neutrino sources, with the diffuse neutrino flux
$I_{\nu_i}(E)$ given by Eq.(\ref{nu-flux}). 
Using the formalism developed in Section V, one can write down the flux
of $Z^0$-bosons with resonant energy $E_0$:
$$
I_{Z^0}=2\xi \sigma_t\delta_{\nu}n_{\nu_i}R_{LSS}I_{\nu_i}(E_0)E_0.
$$
Assuming a power-law spectrum of hadrons in $Z^0 \to hadrons$ decay 
and using Eq.(\ref{nu-flux}) for $I_{\nu_i}(E_0)$, one 
can easily calculate the UHE photon flux from the halo or LG:
\beq
I_{\gamma}^{LLS}(E)=k_{\gamma}\xi\zeta \sigma_t\delta_{\nu}n_{\nu_i}R_{LSS}
P_{osc}\frac{c}{4\pi t_0^2}\left( \frac{E}{m_X}\right)^{-p},
\label{UHE-ga}
\eeq
where $R_{LSS}$ is $\sim 1~Mpc$ is the case of LG, and $\sim 100~kpc$
in the case of galactic halo, $\zeta=r^2\mu/m_X^2$, $\xi=\cos^2\theta,\;\; 
\sin^2\theta$ or 1 and $k_{\gamma}$ is given by
\beq
k_{\gamma}= \frac{4 (2-p)^2 2^{-2p}}{p(p-1)}\frac{\Gamma_{had}}{\Gamma_{tot}}
\eeq
where $\Gamma_{had}/\Gamma_{tot}$ is the ratio  of $Z^0$ decay widths,
equal to 0.7. For $p=1.5 ~~ k_{\gamma}=0.116$ and for $p=1.3 ~~ 
k_{\gamma}=0.58$.

With $\delta_{\nu}=\delta_{\nu}^{max} \sim m_{\nu}^3$, the
UHE gamma-ray flux given by Eq.(\ref{UHE-ga}) is proportional to 
$m_{\nu}^3$ and $m_X^p$. For a fixed overdensity, the flux does not depend 
on the neutrino mass and depends only on the mass of X-particle as $m_X^p$.

As a numerical example let us consider the case of a gamma-ray flux from LG 
with two neutrino flavors and  with a 
degenerate mass $m_{\nu}=2~eV$ ($\xi=1$), 
taking the mass of 
X-particle $m_X=1\cdot 10^{15}~GeV$ and $\zeta=\zeta_{max}=10$. For
$p=1.5$ and $p=1.3$, the values of 
$E^3I_{\gamma}(E)$ at $E=1\cdot 10^{20}~eV$ are equal to 
$2.3\cdot 10^{24}~eV^2/m^2 s sr$ and $1.8\cdot 10^{24}~eV^2/m^2 s sr$, 
respectively, i.e. close to the observed values. 

It is interesting to derive an upper limit for the UHE gamma-ray flux
inside a LSS, using e-m cascade production in the space outside it. 
For LSS with a linear size $R_{LSS}$ and neutrino overdensity $\delta_{\nu}$,
one obtains

\beq
I_{\gamma}^{max}(E)=\frac{2}{3}(2-p)
\frac{\delta_{\nu}R_{LSS}}{ct_0}\frac{c}{4\pi}\frac{\omega_{cas}}{E_0^2}
\left( \frac{E}{E_0}\right)^{-p}.
\label{max-flux}
\eeq
From Eq.(\ref{max-flux}) one can see that the upper limit does not
depend on $m_X$ and is proportional to $m_{\nu}^{5-p}$ for
$\delta_{\nu}=\delta_{\nu}^{max}$. 
For the parameters
of LG ($R_{LG}=1~Mpc$ and $\delta_{\nu}^{LG}=13 m_1^3$) and 
$\omega_{cas}=1\cdot 10^{-6}~eV/cm^3$ one obtains, at  
$E=1\cdot 10^{20}~eV$, fluxes equal to  
$E^3I_{\gamma}(E)=5.0\cdot 10^{23}m_1^{3.5}~eV^2/m^2 s sr$ and 
$2.4\cdot 10^{23}m_1^{3.7} eV^2/m^2 s sr$ for $p=1.5$ and $p=1.3$, 
respectively. At $m_{\nu} > 2~eV$ both upper limits are consistent
with observations.

Turning the argument around, one can obtain a lower limit on the 
neutrino mass
from the condition $I_{\gamma}^{max}(E)>I_{obs}(E)$ at $E=1\cdot 10^{20}~eV$:
$m_{\nu} \gtrsim 2~eV$.

More accurate calculations with realistic QCD spectra from $Z^0$ decay
are given in the accompanying paper \cite{BBV}

\section{Discussion and conclusions}

Mirror matter is a natural option in models with $G\times G'$ symmetry,
in particular in superstring models $E_8\times E_8'$. 
The coupling constants, the Higgs vev's, and the symmetry breaking
patterns in the two sectors may or may
not be the same, depending on whether or not the discrete $P$-symmetry
interchanging the sectors is spontaneously broken.



We assume that the two 
sectors communicate due to a non-renormalizable interaction
(\ref{d5}), where the case $\Lambda\sim m_{pl}$ corresponds to gravitational 
interaction. These interactions result in neutrino masses and neutrino 
oscillations,
including the oscillations between ordinary and mirror (sterile) 
neutrinos. 

Cosmological restrictions rule out a wide class of 
hidden-sector models; they are particularly severe for the symmetric models 
in which the $P$-symmetry is unbroken. In such
models, the number of light particles is
doubled, and this is excluded by the cosmological nucleosynthesis. 
The nucleosynthesis 
constraints can be avoided by suppressing the temperature of the
mirror matter, accompanied by an introduction of a lepton 
asymmetry, which suppresses $\nu \to \nu'$ oscillation.  

Inflationary scenarios resulting in different temperatures in the
ordinary and mirror sectors can easily be constructed for asymmetric
HS models.  In the case of symmetric models, we discussed a
two-inflaton scenario, first outlined in \cite{KST}.  
The inflatons $\phi$ and $\phi'$ belong to the ordinary and hidden
sector, respectively.  In regions of space where
$\phi'$ reaches the minimum of its potential earlier than $\phi$,
the products of $\phi'$ 
decay are diluted by the expansion driven by the ordinary 
inflaton $\phi$. When $\phi$ also rolls to the bottom of its potential,
we get a superhorizon region
dominated by ordinary matter. In stochastic inflation, superbubbles 
dominated by mirror matter are equaly often produced. 
We have shown that this scenario can work only with a suitable choice
of the inflaton potential: for some choices the slow rolls of the two
inflatons get synchronized, resulting in equal densities of mirror and
ordinary matter.  We also suggested an alternative version of the
two-inflaton model where the potential has a metastable minimum.  Then
the inflating false vacuum decays through nucleation of $\phi$- and
$\phi'$-bubbles, and
the segregation of ordinary and mirror matter is achieved in a natural
way.

Despite the suppression of mirror matter, hidden-sector 
topological defects can 
dominate over the ordinary ones. Once again, this can be easily
arranged in asymmetric HS models.  In symmetric models, the two
sectors have the same types of defects with identical properties, but
the cosmological densities of the defects need not be the same.  
We illustrated this possibility by a two-inflaton model with a 
curvature-driven phase transition (see section III). 
In this model, HS
topological defects are produced in a phase transition during
inflation, when the mirror inflaton $\phi'$ is already at the minimum of its
potential.  The phase transition is triggered when the spacetime
curvature (which is driven by the ordinary inflaton potential) 
decreases to some critical value. If this happens sufficiently close
to the end of inflation, the resulting defects are not
inflated away.  
The corresponding phase transition in the ordinary matter occurs much 
earlier, and ordinary topological defects are completely 
diluted by inflation.
Thus, in the two-inflaton scenario we can have a desirable situation when 
the universe is dominated by ordinary matter and hidden-sector
topological defects.

HS topological defects produce high-energy neutrinos in the chain of 
decays of 
superheavy particles -- constituent fields of the defects. All decay 
products are invisible in the ordinary matter and only mirror neutrinos 
oscillate into the ordinary world. 
The flux of neutrinos from ordinary topological defects is limited by 
cascade photons which are produced in the same 
decays of pions as neutrinos. This restriction is absent in the case of HS
defects. However, the cascade limit for mirror neutrinos reappears, 
though in a weaker form. After $\nu' \to \nu$ oscillation, ordinary 
neutrinos produce hadrons, $e^+e^- \;\; \mu^+\mu^-$, and $\tau^+\tau^-$ 
in the resonant scattering off the background (dark matter) neutrinos:
$\bar{\nu}+\nu_{DM} \to Z^0 \to hadrons$, or $l^+l^-$. These particles 
(or products of their decay) initiate electromagnetic cascades on the 
microwave photons. The smaller is the mass of DM neutrinos, the
stronger is the cascade 
upper limit [see Eq.(\ref{lim2})]. This is because the resonant energy   
is inversely proportional to the neutrino mass: $E_0=m_Z^2/2m_{\nu}$. In QCD 
spectra, most of the energy is carried by high energy particles, and thus 
more energy is transferred to the e-m cascade when $E_0$ is large.

As a specific example of mirror topological defects, sources of high energy 
neutrinos, we studied the necklaces -- magnetic monopoles connected by 
strings, with each monopole being attached to two strings. We found
that, for a reasonable choice of model 
parameters, the diffuse neutrino flux can be three orders of magnitude 
higher than that 
from ordinary necklaces, being still consistent with the cascade  
upper limit imposed by the 
resonant production of $Z^0$ bosons. Note, however, that the accuracy 
of our calculations is limited by the power-law approximation of the energy 
spectra.

A diffuse flux of 
UHE neutrinos from HS topological defects can produce the 
observed flux of UHE cosmic rays due to resonant interaction with the Dark  
Matter neutrinos in the Local Group, if the mass of the target
neutrino is 
$m_{\nu}> 2~eV$. The atmospheric-shower producing particles in this case 
are UHE photons. Their spectrum does not exibit a cutoff, 
because of the relatively small size ($R \sim 1~Mpc$) of LG. 

This model of UHE cosmic rays requires
mass degenerate neutrinos with $m_{\nu} >2~eV$. Our model for neutrino masses 
and mixing is described in the Appendix. Mirror muon neutrino $\nu'_{\mu}$ is 
maximally mixed with the ordinary $\nu_{\mu}$ neutrino, and both have  
masses $m_{\nu} \sim 2~eV$. Their mass difference, 
$\Delta m^2 \approx 2\cdot 10^{-3}~eV^2$, is responsible for the atmospheric 
neutrino oscillations observed in Super-Kamiokande. Solar neutrino anomaly
is explained by $\nu_e \to \nu_{\mu}$ LMA MSW solution with 
$\Delta m^2 \approx 4\cdot 10^{-5}~eV^2$ and $\sin^2 2\theta \approx
0.80$. Thus, we assume that the three 
neutrinos, $\nu_e,\;\; \nu_{\mu}$ and $\nu'_{\mu}$ are maximally mixed and 
mass degenerate ($m_{\nu} \sim 2~eV$).

The calculated neutrino flux is below the upper limits obtained from 
horizontal air shower observations at EAS TOP \cite{EASTOP} and 
AGASA \cite{AGASA} at $10^6 - 10^7~GeV$ and marginally below 
Fly's Eye limit \cite{FE} at $10^{11}~GeV$. The predicted neutrino
fluxes can be detected by this technique with the 
help of these and future bigger arrays, like 
e.g. ``Auger'' detector \cite{Auger}. However, the best hope for detecting
these neutrinos probably rests with the future satellite detectors such as 
OWL (Orbiting Wide Field Light Collector) \cite {OWL} and AIRWATCH 
\cite{AIRWATCH}.\\*[2mm]
{\bf ACKNOWLEDGEMENTS}\\*[2mm]
We thank Alexei Smirnov for discussion, helpful comments and
constructive criticism. The participation of Pasquale Blasi in the
part of this work (in preparation) is gratefully acknowledged. We are
grateful to  Zurab Berezhiani, Alexander Dolgov, Gia Dvali, Anjan
Joshipura and Francesco Vissani for useful
discussions.  The work of A.V. was supported in part by the National
Science Foundation.

\appendix

 \section{Mass degenerate neutrinos in mirror models}

We consider $G\times G'$ model with EW symmetry $SU_2(L)\times U_1$ in
$G$ and $SU'_2(R)\times U'_1$ in $G'$. $G$ and $G'$ representations communicate
through operators of dimension $d=5$ with a scale $\Lambda < m_{pl}$.

The particle content of the EW group relevant to the neutrino masses is
\beq
\psi_L= \left( \begin{array}{c}
\nu_L\\ l_L \end{array} \right) ,~~~ l_R,~~~
\phi=\left( \begin{array}{c}
\phi_+\\ \phi_0 \end{array} \right),~~~
\phi^c=\left( \begin{array}{c}
\phi_0^*\\ -\phi^*_+ \end{array} \right)
\label{partL}
\eeq
for $SU_2(L)\times U_1$, and
\beq
\psi'_R= \left( \ba {c}
\nu'_R\\ l'_R \ea \right),~~~ l'_L,~~~
\phi'=\left( \ba {c}
\phi'_+\\ \phi'_0 \ea \right),~~~
\phi'^c=\left( \ba {c}
\phi'^*_0\\ -\phi'^*_+ \ea \right)
\label{partR}
\eeq
for $SU'_2(R)\times U'_1$. Here, $\phi$ and $\phi'$ are the Higgs fields and 
$l$ and $l'$ are charged leptons.

There are no light singlets $\nu_R$ and $\nu'_L$ in our model. There 
are 4 neutrino
states: $\nu_L,\;\nu_L^c,\; \nu'_R, \; \nu'^c_R$, in terms of which the
most general expression for the mass matrix is
\beq
{\cal L}\sim  \left( \bar{\nu}_L~~\bar{\nu'}^c_R \right)
\left( \ba {cc}
m_L & M\\
M   &  m_R \ea \right)
\left( \ba {c}
\nu_L^c\\ \nu'_R \ea \right) +h.c.
\label{matrix}
\eeq
Since $\bar{\nu'}^c_R \nu^c_L=\bar{\nu}_L \nu'_R$, there are three independent
mass operators in Eq.(\ref{matrix}): $\bar{\nu}_L\nu'_R, \;\; 
\bar{\nu}_L \nu_L^c, \;\; \bar{\nu'}^c_R \nu'_R$, and they are generated
in our model. Indeed, these operators are:
\begin{eqnarray}
SU_2(L)\times U_1:~~ \left(\bar{\psi}_L \phi^c\right)
\left( \phi^c \psi^c_L\right) \rightarrow \bar{\nu}_L \nu_L^c
\label{intra}\\
SU'_2(R)\times U'_1:~~ \left( \bar{\psi}'_R \phi'^c\right)
\left( \phi'^c \psi'^c_R\right) \rightarrow \bar{\nu}'_R \nu'^c_R
\label{intraR}\\
intergroup: ~~ \left(\bar{\psi}_L\phi^c\right)
\left( \phi'^{c+} \psi'_R\right) \rightarrow \bar{\nu}_L \nu'_R
\label{inter}
\end{eqnarray}
Arrows show the neutrino mass operators after the EW symmetry breaking,
$<\phi_0>=v_{EW}$ and $<\phi'_0>=v'_{EW}$. Communication of visible
and mirror sectors is accomplished by the {\em intergroup} term in 
Eq.(\ref{inter}). It has a dimensional scale $\Lambda$, and thus 
one obtains 
\beq 
M = v_{EW}v'_{EW}/\Lambda.
\eeq
This is the basic neutrino mass scale in our model, and we want it
to be $M \sim 1~eV$. For $v'_{EW}/v_{EW} \sim 10$, we need 
$\Lambda \sim 10^{14}~GeV$.

The scales of d=5 terms operating inside $SU_2(L)\times U_1$ and   
$SU'_2(R)\times U'_1$ groups, $\Lambda_L$ and $\Lambda_R$, can be different. 
For our model (see below) we need the following hierachy of masses:
\beq 
m_L < m_R \ll M.
\label{hier}
\eeq
It can be provided by $\Lambda \ll \Lambda_R < \Lambda_L$, or in the 
model-dependent way. One can observe, for example, that both intragroup
terms (\ref{intra}) and (\ref{intraR}) violate the lepton number defined for
the doublets as 
$L_{\psi}=L_{\psi'}=1$, while the intergroup operator (\ref{inter})
conserves it. One can
build a model with one universal $\Lambda$ where the intragroup d=5 
operators are forbidden and thus $m_L$ and $m_R$ are suppressed.

Let us assume a local $\tilde{U}_1$ symmetry for massless particles before 
symmetry breaking, with the following charge
assignment: $q=+1$ for $\psi_L,\;\; l_R,\;\;\psi'_R,\;\; l'_L$ and 
$q=0$  for $\phi,\;\;\phi'$.

The terms (\ref{intra}) and (\ref{intraR}) do not conserve $q$. Let us
introduce two new scalar $SU_2$ singlets, $\Phi$ and $\Phi'$, with 
$q= 2$. Now, apart from the operator (\ref{inter}), we can write two other 
$SU_2$-singlet operators conserving the charge $q$ and the electric charge.
They are the 
following $d=6$ operators: 
\beq     
\frac{\Phi}{\Lambda^2}\left(\bar{\psi}_L \phi^c\right)
\left( \phi^c \psi^c_L\right),~~~ \frac{\Phi'}{\Lambda^2}
\left( \bar{\psi}'_R \phi'^c\right)\left( \phi'^c \psi'^c_R\right).
\label{d=6}
\eeq
After EW and $\tilde{U}_1$ symmetry breaking with vev's $<\Phi>=V$ and 
$<\Phi'>=V'$, respectively, one obtains 
\beq
M=v_{EW}v'_{EW}/\Lambda,~ m_R=v_{EW}^{'2}V'/\Lambda^2,~ 
m_L=v_{EW}^2V/\Lambda^2,
\eeq
which satisfy the hierarchy (\ref{hier}).

Let us now come back to the mass matrix
(\ref{matrix}). Its diagonalization gives the masses of eigenstates
and the mixing angle for visible and mirror neutrinos:
\begin{eqnarray}
m_{1,2}&=& \left( m_R+m_L \pm \sqrt{4M^2+(m_R-m_L)^2}\right) \approx 2M\\
\label{eigmass}
\Delta m^2&=&m_2^2-m_1^2 \approx 2(m_R+m_L)M\\
\label{delta}
\sin 2\theta&=&\frac{2M}{\sqrt{4M^2+(m_R+m_L)^2}} \approx 1
\label{mix}
\end{eqnarray}
The hierarchy of masses (\ref{hier}) provides mass degenerate neutrinos 
with (almost) maximal mixing.

Till now we considered asymmetric mirror models. In the case of 
symmetric
models \cite{FoVo}, the masses $M,\;\;m_L$ and $m_R$ are considered as
free parameters, and thus the hierarchy condition (\ref{hier}) can be 
arbitrarily fulfilled.

As a realistic example we can consider the case when $\nu_{\mu} \to \nu'_{\mu}$
oscillations explain the atmospheric neutrino anomaly 
($\Delta m^2 \approx 2\cdot 10^{-3}~eV^2$) and LMA MSW explains the
solar neutrino problem ($\nu_e \to \nu_{\mu}$ oscillation). In this
case, $2M\sim 1~eV$ and $m_R \sim 1\cdot 10^{-3}~eV$. All three
neutrinos, $\nu_{\mu},\; \nu_e$ and $\nu'_{\mu}$, must have 
mass $2M \sim 1~eV$ and be bimaximally mixed. The oscillation
length ($\nu'_{\mu} \to \nu_{\mu}$) for HE neutrinos with resonant energy
$E_0$ is only $l_{osc} =4\pi E_0/\Delta m^2 = 5\cdot 10^{20}~cm$.
The values of $\Delta m^2$ in this case are $2\cdot 10^{-3}~eV^2$
and $\sim 4\cdot 10^{-5}~eV^2$ for the atmospheric and solar neutrinos, 
respectively; they are not affected much by radiative corrections 
\cite{rad}. Radiative splitting affects mostly $\tau$ neutrino,
because of the large Higgs coupling with the tau. This effect is not
important in our scenario where only three other neutrinos,
$\nu_{\mu},~~ \nu'_{\mu}$ and $\nu_{\tau}$, are degenerate in mass.

\end{document}